\documentclass[english,twocolumn,prb,superscriptaddress,showpacs, amsmath]{revtex4-2}
\usepackage[utf8]{inputenc}
\usepackage{amsmath,amssymb,bm,graphicx,color}
\usepackage{varwidth}
\usepackage[bookmarks=false]{hyperref}
\usepackage[usenames,dvipsnames]{xcolor}
\usepackage[super]{nth}
\usepackage{ulem,lipsum}
\usepackage{subcaption} 
\usepackage[export]{adjustbox} 
\usepackage{cleveref} 
\usepackage{tikz}
\usetikzlibrary{shapes}
\usepackage{soul}
\usepackage{braket}
\usepackage{etoolbox}

\makeatletter
\def\@seccntformat#1{\csname the#1\endcsname\quad}
\makeatother

\begin{document}

\title{Revealing quantum geometry effects in magic angle twisted bilayer graphene using the circular photogalvanic effect}

\author{Eylon Persky}
\affiliation{Geballe Laboratory for Advanced Materials, Stanford University, Stanford, CA 94305, USA}
\affiliation{Stanford Institute for Materials and Energy Sciences, SLAC National Accelerator Laboratory, 2575 Sand Hill Road, Menlo Park, CA 94025, USA}
\affiliation{Department of Applied Physics, Stanford University, Stanford, CA 94305, USA}
\author{Léonie Parisot}
\affiliation{Geballe Laboratory for Advanced Materials, Stanford University, Stanford, CA 94305, USA}
\author{Minhao He}
\affiliation{Department of Physics, University of Washington, Seattle, Washington, 98195, USA}
\author{Jiaqi Cai}
\affiliation{Department of Physics, University of Washington, Seattle, Washington, 98195, USA}
\author{Takashi Taniguchi}
\affiliation{Research Center for Materials Nanoarchitectonics, National Institute for Materials Science, 1-1 Namiki, Tsukuba 305-0044, Japan}
\author{Kenji Watanabe}
\affiliation{Research Center for Electronic and Optical Materials, National Institute for Materials Science, 1-1 Namiki, Tsukuba 305-0044, Japan}
\author{Pierre A. Pantale\'on}
\affiliation{Imdea Nanoscience, Faraday 9, 28047 Madrid, Spain}
\author{Francisco Guinea}
\affiliation{Imdea Nanoscience, Faraday 9, 28047 Madrid, Spain}
\affiliation{Donostia International Physics Center, Paseo Manuel de Lardiz\'abal 4, 20018 San Sebastián, Spain}
\author{Xiaodong Xu}
\affiliation{Department of Physics, University of Washington, Seattle, Washington, 98195, USA}
\author{Aharon Kapitulnik}
\email{aharonk@stanford.edu}
\affiliation{Geballe Laboratory for Advanced Materials, Stanford University, Stanford, CA 94305, USA}
\affiliation{Stanford Institute for Materials and Energy Sciences, SLAC National Accelerator Laboratory, 2575 Sand Hill Road, Menlo Park, CA 94025, USA}
\affiliation{Department of Applied Physics, Stanford University, Stanford, CA 94305, USA}
\affiliation{Department of Physics, Stanford University, Stanford, CA 94305, USA}

\begin{abstract}

Magic-angle twisted bilayer graphene (MATBG) hosts flat bands whose correlated and topological phases depend sensitively on the electronic point-group symmetry, which is set by the device configuration including twist angle, heterostrain, and substrate and cap-layer coupling. Here we report near-infrared photocurrent measurements of a MATBG device encapsulated by an hBN substrate and a WSe$_2$ cap layer, at a twist angle of 1.05°. Using spatially resolved photocurrent imaging with polarization analysis, we disentangle the photothermoelectric response from the circular photogalvanic effect (CPGE) across all fillings of the moir\'e flat bands. A finite CPGE at normal incidence, forbidden in high-symmetry structures, directly constrains the point-group symmetry of MATBG and reveals a Berry curvature dipole intrinsic to the MATBG flat bands, activated by the symmetry lowering from heterostrain and the asymmetric hBN/WSe$_2$ environment. Our results establish nonlinear photocurrent as an all-optical probe of symmetry and quantum geometry in proximity-engineered moir\'e materials.

\end{abstract}

\date{\today}

\maketitle
%\section{Introduction}

\noindent \textbf{I. Introduction:} 

\vspace{2mm}
Nonlinear optical techniques present indispensable access to the electronic structure of quantum materials, offering unique access to symmetry properties that are often obscured in conventional transport measurements. In particular, the circular photogalvanic effect (CPGE), where circularly polarized light drives a dc current that reverses with helicity, is an exquisitely sensitive probe to the interplay between quantum geometry, strong correlations, and broken symmetries in quantum materials. Thus, CPGE experiments can uncover several frontiers of condensed-matter physics, particularly associated with two-dimensional (2D) heterostructures such as twist-stacked moir\'e materials.

Focusing on Magic Angle Twisted Bilayer Graphene, CPGE provides a relatively direct probe of valley polarization, Berry curvature, and chiral optical selection rules - quantities that are otherwise hard to access. First, the flat bands and strong interactions present in MATBG are expected to amplify nonlinear optical responses like CPGE, enabling stronger photocurrents at lower optical power than in more conventional 2D materials. Since CPGE is forbidden in centrosymmetric systems, it is a sensitive probe to study many of the symmetry-broken electronic states in MATBG devices, including spontaneous inversion symmetry breaking, nematic order, valley polarization effects and time-reversal symmetry breaking, all can be mapped as function of density, temperature, and displacement field.  CPGE couples directly to Berry curvature distribution in momentum space and thus provides experimental access to valley-selective excitation, Berry curvature dipole moments as well as to orbital magnetic moments. It can therefore reveal the orbital and real-space character of wave-functions, helping discriminate between competing theoretical models of MATBG's insulating and superconducting phases  adding information that complements but goes beyond what conventional transport can provide  \cite{Cao2018correlated,Cao2018unconventional}\\

\noindent \textbf{II. Summary of Results:}

\vspace{2mm}
\indent In this paper we present results of normal-incidence CPGE measurements using 1550 nm light, on a previously studied \cite{He2025,Persky2026} WSe$_2$-capped MATBG device \cite{Polski2022} 
as a function of band-filling.  Our studies reveal a strong correspondence between the CPGE data, the moir\'e filling, and a close tracking of the Berry curvature dipole calculated around the Fermi surface.  These findings strongly suggest substantial contribution from the moir\'e electronic structure and the associated quantum geometric properties, which is further rationalized by the high cleanliness of the sample leading to long thermalization time (see CPGE in the Supplemental Material \cite{Supp}). In our experiments a superposition of pure right and left circularly polarized light, with relative intensity that can be continuously dialed between the two circular polarization states, shines on the device and the photo-generated current and voltage are simultaneously recorded. The CPGE data is established after careful subtraction of the photo-thermoelectric (PTE) current, whose behavior is characterized independently.  

\begin{figure}[ht!]
\centering
\includegraphics[width=1.0 \columnwidth]{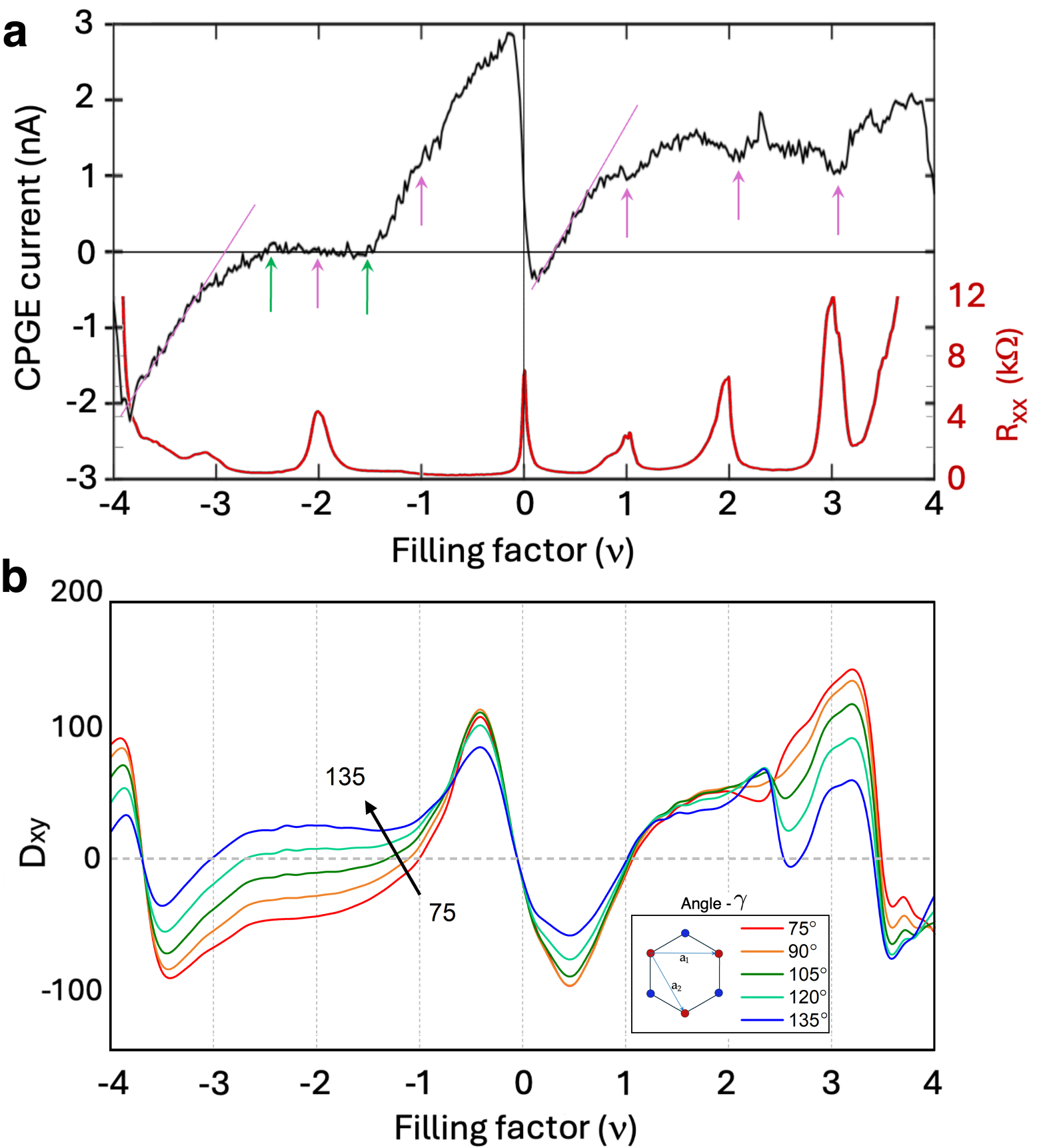}
\caption{ (a) Measured CPGE current (top) and  longitudinal resistance (bottom) vs. band filling.   Arrows and straight lines point to the key features of correlated states in the band. (b) Calculation of Berry-Curvature Dipole (BCD) with a uniaxial 0.3 \% heterostrain along the K--K$^\prime$ direction. Calculation parameters:  Twist-angle, $\theta=1.05^{\circ}$, sublattice potential $\Delta_m=10$ meV, interlayer moir potentials $u=$ 79.7 meV and $u^\prime=$ 97.5 meV, $\hbar v_F= -2.13 a$ with $a=0.246 $ nm, dielectric constant $\varepsilon=6$ and double-gate distance $d_g=$ 40 nm. Legend shows the BCD orientation (angle-$\gamma$) with respect to the $\textbf{a}_1$ graphene lattice vector.}
\label{cpgeX}
\end{figure}

Our key  experimental results are:  
{\it i}) A robust finite CPGE is found over a wide range of electron densities and temperatures from $\sim10$ K down to 0.3 K, suggesting that the device's point group symmetry is lowered to \textit{C}$_1$ at the heterostructure level.   
{\it ii}) Low-temperatures PTE voltage data in the same temperature range exhibit features at the integer fillings of $\nu= +1, \pm 2$ and $+$3, as well as at the charge neutrality point (CNP), that become increasingly sharper with lower temperatures, where except for an overall bias that increases with filling, resemble a behavior expected from Mott's semiclassical relation (MSR) \cite{Cutler1969}. 
{\it iii}) An offsetted CPGE sign change near $\nu=0$ highlights a strong particle-hole asymmetry. While at 1550 nm we might expect the interband effects to dominate, we do observe highly discernible features which are clearly associated with Fermi surface effects as we show in the comparison of the data to the longitudinal electrical resistivity of the device (Fig.~\ref{cpgeX}a).
 {\it iv}) Remarkably, at a range of fillings between $\nu =-2.5$ and $\nu = -1.5$, coincidental with the PTE-Voltage deep region around $\nu=-2$, the CPGE vanishes to very high precision.  
 {\it v}) The zero-CPGE emerges as a sharp onset at $\nu=-1.5$  which remains sharp even at 10 K, suggesting a symmetry breaking effect associated with an energy scale $> $ a few meV. The zero CPGE plateau disappears with a softer, temperature dependent manner below $\sim-2.5$. 

The above observations correspond more to Fermi-surface tracking of the CPGE, where Berry curvature dipole should dominate, rather than to  interband transitions. Indeed, Pantaleón {\it et al.}, \cite{Pantaleon2022}  pointed out that a combination of a small uniaxial heterostrain,  hexagonal boron nitride (hBN) substrate, and electron-electron interactions treated within a self-consistent Hartree formalism are enough to explain the emergence of strong electron-hole asymmetry and a large Berry Curvature Dipole (BCD) that appears when the chemical potential is tuned to a spectral region entangled with the remote bands. Their calculations largely explained a recent experiment that observed nonlinear Hall currents in MATBG on hBN \cite{Duan2022}. In Fig.~\ref{cpgeX} we show calculation of Berry curvature dipole, where by choosing a reasonable set of device parameters, we find that the band-filling dependence of the CPGE is qualitatively consistent with Berry curvature dipole (BCD) calculations in the presence of strain for negative fillings, while for positive fillings there is a stronger background CPGE that takes the data all positive, yet vestiges of the correlation anomalies can still be observed. While the experimentally measured CPGE vanishes within the interval $-2.5<\nu<-1.5$, the present Berry-curvature-dipole calculation already produces a very small response in the same filling range, although not an exactly zero signal. The calculation also shows a sign change and a strong inflection centered near $\nu\simeq -2$ for a broad range of effective dipole orientations. This suggests that the observed zero-CPGE plateau is connected to the same Fermi-surface geometric mechanism captured by the Berry-curvature dipole calculation. In the experiment, additional  correlations not included in the present Hartree treatment may further stabilize this near-zero response and lock the photocurrent to zero (for further discussion on the calculations, see \cite{Supp}).\\

\noindent \textbf{III. Experiment and Analysis:}

\begin{figure}[ht!]
\centering
\includegraphics[width=1.0 \columnwidth]{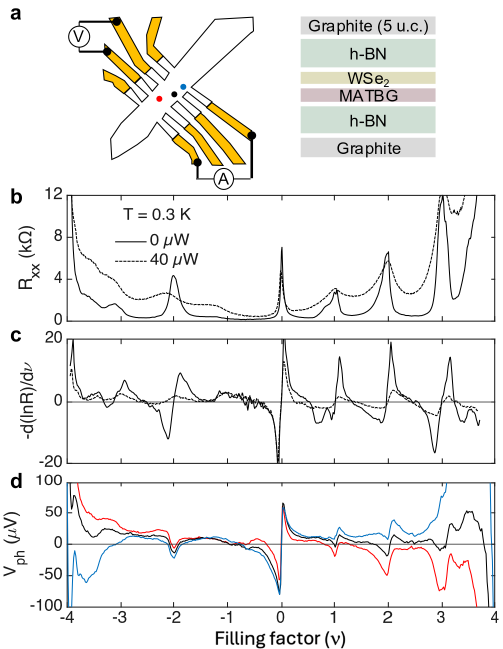}
\caption{(a) Schematic of the device configuration. For photocurrent measurement, the device is shorted in series with a current amplifier, which measures the current that is induced by the light. The left side shows the van der Waals stack of our dual-gated sample. (b) Longitudinal resistance as a function of carrier density taken at $T = 0.3$ K with and without unpolarized light (incident power, 40 $\mu$W). (c)  Log-derivative of the resistance vs. filling as a test to Mott's semiclassical relation~\ref{eq-MottFormula}, with and without illumination as in (a).  (d) Photovoltage measured at 0.3 K with the center of the beam at the three points colored: blue, black and red. }
\label{fig-Mott}
\end{figure}

\vspace{2mm}
\indent We studied a dual-gated MATBG/WSe$_2$ heterostructure (Fig. \ref{fig-Mott}a), where the top gate was fabricated from a thin (3-5 layers) graphite flake to allow optical measurements. The measurement setup is described in details in the Supplemental Material \cite{Supp}. Broadband NIR light launched into a $^3$He cryostat was converted to left (right) circularly polarized light before interacting with the sample. Controlling the ratio of left/right circularly polarized light incident on the sample, the respective photo-current and photo-voltage were measured using lock-in amplifier. The resistance of the sample was measured separately both in Dark and illuminated conditions, by biasing the sample with a small ac current (1-10 nA). Unless otherwise specified, all data were taken in a zero displacement field, following previous tests that transport properties and optical responses did not show any noticeable displacement field dependence \cite{Persky2026}.

Fig. \ref{fig-Mott}b shows $R_{xx}$ measured at 300 mK with and without illumination, as a function of filling (in electrons per moir\'e unit cell).  As we previously reported on the same device \cite{Persky2026}, the transport shows characteristic insulating peaks at integer filling $+1, \pm2, \pm3$, which are broadened upon illumination due to light-induced heating (about 2 K increase in the effective electron temperature), without noticeable photodoping.  Thus, we first characterize the contribution of thermoelectric effects to the photo-transport by measuring photocurrents and photovoltages generated by unpolarized light (Fig. \ref{fig-Mott}d), where polarization-dependent contributions, such as the PGE or geometry-related effects \cite{Castilla2019,Candussio2020,Semkin2023}, should largely cancel out compared to the leading thermoelectric effect contributions.  

Measuring simultaneously photovoltage and photocurrent, the deduced Seebeck coefficient ($S=\Delta V/\Delta T$) will only be affected by the high resistance  near integer fillings, which we further discuss below. Examination of the measured photovoltage (Fig. \ref{fig-Mott}d), we observe sharp saw-tooth-like features at charge neutrality, and positive integer fillings, with a more symmetric deep near $\nu=-2$, roughly similar to previous extraction of thermoelectric voltage in near-magic angle bilayer graphene devices \cite{Ghawri2022, Merino2025}. However, performing a light-illuminated gate-defined MATBG p-n junctions experiment, Merino {\it et al.} \cite{Merino2025} observed low-temperature sign-preserving of the full sawtooth effect in their inferred Seebeck coefficient, which evolves towards conventional sign-changing thermoelectricity at integer fillings when the temperature increases. To test this effect in our device we positioned the center of the beam at different points (Fig. \ref{fig-Mott}a), observing that at positive fillings the overall sawtooth-response remains the same, becoming sharper at lower temperatures with a sign preference that can be either positive, negative or zero, depending on the position of the beam (Fig. \ref{fig-Mott}d). Since the sign bias is small and not fixed in sign, it can be naturally explained by contact effects or the small contribution a finite current drawn as described above.  Where sign change is roughly symmetric around zero, the photovoltage shows remarkable agreement with the Mott's semiclassical relation \cite{Cutler1969}:

\begin{equation}
    S = -\frac{\pi^2}{3}\frac{k_B^2T}{e}\frac{\partial \ln \sigma(\mu)}{\partial\mu} \propto \frac{\partial \ln R}{\partial \nu} \frac{\partial \nu}{\partial \mu},
    \label{eq-MottFormula}
\end{equation}
which we show in Fig. \ref{fig-Mott}c. Here $k_B$ is Boltzmann's constant, $T$ is the temperature, $e$ the electron charge, $\sigma$ is the conductivity, $\mu$ is the chemical potential, $R$ is the resistance and $\nu$ is the band filling, proportional to the carrier density.

\begin{figure*}[ht!]
\includegraphics[width=2.0 \columnwidth]{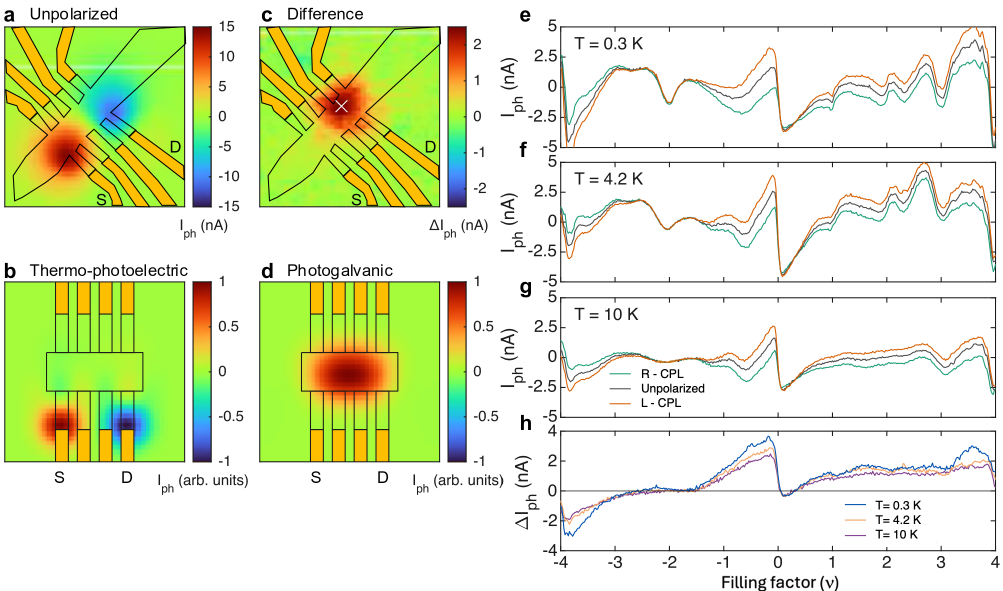}
\caption{(a) Experimental photocurrent map taken at filling $\nu = -0.5$ unpolarized light. The device is overlaid on the image based on a simultaneously acquired reflectivity map, and the source and drain electrodes are indicated. (b) Calculated photocurrent map for a Hall bar geometry assuming the photo-induced current originates from local heating through the thermoelectric effect. (c) Difference map of two photocurrent images taken with right- and left-CPL at $\nu = -0.5$. (d) Calculated photocurrent map assuming the photo-induced current originates from a CPGE. (e)-(g) Carrier density dependence of the photocurrent for unpolarized (black) and CPL (red/green). The beam position is indicted by the cross on panel c. (h) Subtracted curves for right and left CPL from panels (e)-(g), corresponding to CPGE.}
\label{fig-cpge}
\end{figure*}

The total photocurrent collected at the electrodes depends strongly on the geometry of the device and position of the heat source \cite{Song2014}. Particularly, when the beam is symmetrically positioned between the source and drain electrodes, the distribution of thermal currents is symmetric, and no net current is collected; the thermoelectric currents are maximized when the temperature gradient is localized near the source and drain electrodes. We tested this by raster-scanning the beam over the device, and recording the photocurrent collected at the current-meter as a function of the beam position (Fig. \ref{fig-cpge}a; taken at $T = 4.2\,$K and at $\nu = -0.5$), revealing that photocurrent generated by unpolarized light was indeed generated when the beam was positioned over the edge where there is a discontinuity between the thermal conductivity of the MATBG and the leads. 

Unlike the thermoelectric current, the  CPGE-generated photocurrents have a preferred direction and therefore a distinct spatial signature. To measure the CPGE, we change the polarization of the incident light, and take photocurrent maps for left and right circularly polarized light (CPL). The difference image $\Delta I_{ph}$ (Fig. \ref{fig-cpge}c) shows a response that is finite only when the light illuminates the MATBG and does not change sign as the beam is moved from the source to the drain electrode. The observations can be reproduced in simulations assuming the induced current flows along the Hall bar (Fig. \ref{fig-cpge}d). Note that only a part of the MATBG device produces this response. The spatial structure of the polarization dependent response rules out a dichroic contribution, where helicity-dependent absorption leads to stronger thermoelectric effect for right or left CPL. In addition, the concentration of the response at the center of the device rules out contribution from the contacts (which in non-ohmic situations typically induce photovoltaic effects \cite{Lee2008,Park2009,Xia2009}), or antenna effects \cite{Castilla2019} or  as the source of the current. We therefore conclude that the measured $\Delta I_{ph}$ is a circular photogalvanic current.

We now turn to study the carrier and temperature dependence of the CPGE. Figure \ref{fig-cpge}f shows the photocurrent measured as a function of the carrier density at $T = 4.2$ K. There is a clear difference between the response to right and left CPL at almost all fillings of the moir{\'e} unit cell, on top of the thermoelectric contributions which dominate the response. To better visualize the CPGE, we plot the difference between the right and left CPL responses in Fig. \ref{fig-cpge}g. The CPG current changes sign at $\nu =0$ and $\nu = 1$  and is relatively filling-independent for $1<\nu<4$. For hole doping, however, the CPGE vanishes for a wide range of fillings $-2.5 < \nu <-1.5$. At 10 K, which is above the AHE temperature for any filling \cite{He2025,Persky2026},  all the correlated insulator resistive and thermoelectric features are smeared (Fig~\ref{fig-cpge}e). However, the CPGE still showed the same overall behavior as in 0.3K, in particular the sharp change in behavior at the neutrality point and the sharp onset of CPGE at $\nu=-1.5$.\\

\noindent \textbf{IV. Discussion:}

\vspace{2mm}
\indent We first analyze the symmetry requirements for CPGE. We may write the DC photocurrent that is generated by an electric field $\mathbf{E}(t)=\text{Re}\{\mathbf{E}e^{i\omega t}\}$ where $\mathbf{E}$ encodes the intensity and polarization of the field, and $\omega$ is the frequency as
\begin{equation}
    J^{(2)}_i=\zeta_{ijk}\left(E_jE_k^* +E_kE_j^* \right) +\eta_{ijk}\left(E_jE_k^* -E_kE_j^* \right),
\end{equation}
where $\zeta$ and $\eta$ encode the linear and circular photogalvanic responses, respectively. Since we measure a 2D sample, the current is in-plane, and normal incidence guarantees $E_z=0$. Since $\eta$ is anti-symmetric $\eta_{ijk} = -\eta_{ikj}$, our experiment probes $\eta_{xxy}$ and $\eta_{yxy}$. The normal incidence configuration therefore restricts the symmetry to the non-centrosymmetric gyrotropic groups \textit{C}$_{1h}$ (also labeled:  $\bar{2}$, ``m'' or \textit{C}$_s$), or \textit{C}$_1$ (also labeled: 1) \cite{Birss1963}. 

While a single sheet of graphene has \textit{D}$_{6h}$ point-group symmetry, the symmetry of twisted bilayer graphene is reduced since an incommensurate twist angle reduces the six-fold rotation to three-fold, and breaks inversion. The resulting point group, \textit{D}$_3$, is still too high to support neither AHE states at integer fillings, which we previously reported on it \cite{He2025,Persky2026}, nor CPGE at normal incidence. Further lowering of the symmetry can arise from ``structural'' origins, such as interaction and alignment to hBN, uniaxial strain, relaxation of atomic positions and twist angle disorder \cite{Uri2020}, or it may be induced by correlations, such as a nematic state \cite{Cao2021} or a charge density wave \cite{Jiang2019}. The fact that the CPGE does not change qualitatively as a function of filling in the range $1\leq\nu\leq4$, and remains present from $0.3$ K to $10$ K, suggests a structural origin rather than a low-temperature instability. We attribute this low symmetry to the combined effect of asymmetric encapsulation by hBN and WSe$_2$, together with induced strain, which lowers the effective point group of the device to \textit{C}$_1$. This is consistent with the Berry-curvature-dipole picture discussed below, since in \textit{C}$_1$ no point-group symmetry forbids an in-plane Berry-curvature dipole. In addition, the self-consistent Hartree reconstruction modifies the bands on an energy scale of order $10$--$60$ meV, much larger than $k_BT$ in the experimental range, so the corresponding band-geometric response is expected to remain
robust from $0.3$ K to $10$ K. 

In this low symmetry, and with our excitation light of 0.8 eV, the remote bands are as important as the flat bands and the most general expression for the circular photogalvanic tensor at normal incidence takes the form:
\begin{equation}
\eta_{ijk}(\omega) = \frac{\pi e^3}{4\hbar^2\omega} \sum_{n,m} \int \frac{d^2k}{(2\pi)^2} f_{nm}(\mathbf{k}) \, \mathcal{T}_{ijk}^{nm}(\mathbf{k}) \, \delta(\omega_{mn} - \omega)
\end{equation}
where $\omega$ is the light frequency, $n$ and $m$ are band indices and the generalized injection tensor $\mathcal{T}_{ijk}^{nm}$ includes three contributions
\begin{equation}
\mathcal{T}_{ijk}^{nm} = \mathcal{T}_{ijk}^{nm,\text{I}} + \mathcal{T}_{ijk}^{nm,\text{BC}} + \mathcal{T}_{ijk}^{nm,\text{shift}}
\end{equation}
which are in order the I$=$standard Injection Current due to asymmetric scattering, proportional to the velocity, BC$=$Berry Curvature Dipole contribution, and shift$=$ shift current, typically associated with the geometric phase. We discuss the details of these terms in the Supplemental Material (\cite{Supp}). If scattering or relaxation processes are important, at least the velocity term will be multiplied by $\tau_{nm}/(1 + \omega^2\tau_{nm}^2)$, with $\tau_{nm}$ depends on the specific relaxation mechanism.

There are primarily two reasons for the moir\'e quantum geometry effects to appear in the CPGE measurements. First, is when Interband interference between flat band and remote band with different Berry curvatures, which represent a hybrid of interband BCD and shift-current mechanism. Such an interband coherence effect depends on interband dephasing time and various dipole matrix elements, with no explicit dependence on recombination or momentum relaxation. While this effect may dominate the positive fillings, it cannot account for the close correspondence between the calculation and the experiment for negative fillings.  The second mechanism assumes that the photoexcited carriers cascade into the flat band and then respond via anomalous velocity, which we call a pumped BCD mechanism. Here 0.8 eV photons excite carriers from occupied bands into remote (graphene-like continuum) bands via interband transitions $n \to m$ with $\hbar\omega_{mn} \sim 0.8$ eV. These hot carriers rapidly relax/cascade down through phonon and e–e scattering within a time-scale much faster than either momentum relaxation or recombination times,  eventually accumulating in the flat band (which lies near the Fermi level). Once in the flat band, the (now long-lived) carriers are described by a correction $\delta f_{n\mathbf{k}}^{(A)}$ to the equilibrium distribution function $f^0_{n\mathbf k}$.  With the fermi level within the flat bands, 
$\delta f_{n\mathbf{k}}^{(A)}\to \delta f_{F\mathbf{k}}^{(A)}$ is practically a  correction to the equilibrium distribution function in the flat band - $f_{F\mathbf{k}}^{(0)}$, while experiencing the associated anomalous (Berry-curvature) velocity:
\begin{equation}
\mathbf{v}_F^{\rm an} = -\frac{e}{\hbar}\,\mathbf{E}(t) \times \Omega_F^z(\mathbf{k})\,\hat{z},
\end{equation}
under the driving optical field $\textbf{E}(t)$. The product of the helicity-odd $\delta f_F^{(A)}$ and the Berry-curvature anomalous velocity yields a DC current proportional to $\Omega_F^z$, or its dipole $\partial_k \Omega_F^z$
\begin{equation}
D^{zl}_{\rm pumped}(\omega)
=
\sum_{n\in F} \int\frac{d^2k}{(2\pi)^2}
\partial_{k_l}\Omega_n^z(\mathbf k)\;
[\delta f^{(A)}_{n\mathbf k}(\omega)]
\end{equation}
which is the same expression as for the equilibrium Berry curvature sampled near the chemical potential, except for $f_{n\mathbf k}\to\delta f^{(A)}_{n\mathbf k}$ with $n\in$F. 

The persistence of the CPGE response to above 10 K also suggests that a Hartree approach, where coulomb interactions are considered at the mean-field level, together with the effect of the encapsulation and  strain, already capture the essence of the observed CPGE through the calculation of the Berry curvature dipole. Indeed, as emphasized at the outset of this paper, there is a remarkable qualitative agreement between such calculations and the observed CPGE, except in the range $-2.5\leq \nu \leq -1.5$, where the CPGE vanishes to a very high degree.  We note however that the Berry-Curvature Dipole calculations do find a sign change centered around $\nu=-2$ with a wide plateau region may indicate that the observed zero-plateau emerges from the same effect, possibly in the presence of strong correlations not included in the present model which pin the plateau at zero (for further discussion on the calculations, see \cite{Supp}).

While the Seebeck coefficient, similar to the longitudinal conductivity, does not involve quantum geometry effects within straightforward semiclassical Boltzmann-Mott approach, we already noted above that the overall bias in the photovoltage, which shifts away the sign change at integer fillings expected from SMR points to a non-Fermi-liquid behavior in that temperature regime (Fig.~\ref{fig-Mott}d). Nevertheless, we do observe that as we lower the temperature towards base temperature, the sawtooth-like response of the PTE response gets sharper and exhibiting a weaker bias, which we interpret as moving towards a Fermi liquid behavior at lower temperatures. At the same time, a Hartree approach with the same fit parameters used to calculate the BCD fails to reproduce the photovoltage results at any temperature (see Supplemental Material \cite{Supp} ). 

The observations above indicate that the Hartree-renormalized Berry-curvature-dipole picture captures the tendency toward a suppressed CPGE near $\nu\simeq -2$, but does not by itself explain either the detailed PTE response or the exact vanishing of the CPGE over a finite filling interval. These features suggest that additional correlation effects, possibly involving a reconstructed electronic state, are relevant. Indeed, correlations at least at the exchange level have been essential to explain the correlated Chern insulator (CCI) states in MATBG, with Chern numbers $|t|=1,2,3$ and moir\'e filling factors $|s|=3,2,1$~\cite{Nuckolls2020,Saito2021,Park2021,Choi2021,Das2021,Wu2021}. For the same MATBG device studied here, He et al.~\cite{He2025} reported a detailed study of CCI states, together with a cascade of symmetry-broken Chern insulator (SBCI) states that appear primarily upon hole doping of the magnetic subbands. These SBCI states were proposed to retain the spin and valley polarization properties of a parent CCI, while further breaking moir\'e translation symmetry through a charge-density wave that
doubles the moir\'e unit-cell area~\cite{He2025}.

In particular, Ref.~\cite{He2025} showed that the SBCI $(-2,-3/2)$ state, which is relevant to the filling range discussed here, terminates at a threshold magnetic field near a distinct state identified as a gapped incommensurate Kekul\'e spiral (IKS). The two states meet at the rational flux ratio $\Phi/\Phi_0=1/4$~\cite{He2025}. In the present experiment, however, the sharp onset of the zero-CPGE regime occurs near $\nu=-1.5$ already at zero magnetic field. This suggests, as a reasonable conjecture, that the zero-field state may be related to the same tendency toward moir\'e unit-cell doubling inferred from the high-field SBCI $(-2,-3/2)$ state. Since the zero-CPGE region is centered near $\nu=-2$, this conjecture is also naturally connected to the van-Hove singularity, where the Fermi level is susceptible to pinning and the charge distribution is especially sensitive to the self-consistent electrostatic potential. STM studies of MATBG on hBN have reported a $\sqrt{3}\times\sqrt{3}$ reconstruction at the graphene atomic scale near the correlated insulator at  $\nu=-2$~\cite{Nuckolls2023}. They also found broken translation and rotational symmetries on the moir\'e scale, with stripe-like features. while these results show the propensity of the flat-band electrons to symmetry breaking effects near $\nu=-2$ filling factor, the resulting broken state may simply indicate a much higher symmetry of the uncapped MATBG samples, possibly exhibiting at least a \textit{C}$_3$ symmetry.

By contrast, the hBN substrate, the WSe$_2$ cap, and the induced strain reduce the structural symmetry of the device to \textit{C}$_1$. In this low-symmetry environment, a unidirectional $1Q$ stripe CDW with a preferred moir\'e $M$-direction is a natural translation-symmetry-breaking instability. Such a state could double the moir\'e unit cell near $\nu=-1.5$, providing a possible zero-field complement to the high-field SBCI $(-2,-3/2)$ state. A related theoretical precedent was recently found in strained graphene superlattices, where enlarged-unit-cell calculations showed that a purely self-consistent Hartree potential can stabilize translation-symmetry-broken charge configurations, without requiring a magnetic field or exchange-driven order~\cite{Andrade2025}.

While the above discussion  focuses on electronic structure near the Fermi level, the optical excitation energy in our experiment is $\hbar\omega\simeq0.8~{\rm eV}$. At this energy, the CPGE is likely to receive substantial contributions from interband injection processes involving remote moir\'e bands, including higher conduction bands, deeper valence bands, and states with appreciable weight in the graphene-like Dirac sectors. Therefore, the exact cancellation of the CPGE should not be interpreted as a purely low-energy flat-band effect. Rather, the unit-cell-doubling scenario should be viewed as a possible additional ingredient to the Berry-curvature-dipole picture discussed above. The Berry-curvature-dipole calculation already produces a nearly vanishing response near $\nu\simeq -2$, suggesting that the dominant tendency toward suppression is associated with the Hartree-renormalized band geometry near the van-Hove singularity. A translation-symmetry-broken state could further reconstruct the bands by folding states at $\mathbf k$ and $\mathbf k+\mathbf Q$, redistributing optical transitions among the folded eigenstates and allowing a stronger cancellation of helicity-dependent contributions and sensitivity to the proximate van-Hove singularity. A more rigorous account of this possible reconstructed state and of its full-band optical response in the present MATBG device  requires further theoretical development and is left for future work.

\bigskip

\textit{Acknowledgments} -- We thank Andrei Bernevig and Steve Kivelson for fruitful discussions. Work at Stanford University was supported by  the U.~S.~Department of Energy (DOE) Office of Basic Energy Science, Division of Materials Science and Engineering at Stanford under contract No.~DE-AC02-76SF00515. EP was partially supported by the Koret Foundation. Work at the University of Washington is supported by NSF MRSEC DMR-2308979.\\

\textit{Data Availability} -- The data that support the findings of this study are available from the corresponding authors upon reasonable request.

\bibliography{References.bib}

\onecolumngrid

\setcounter{section}{0}
\setcounter{figure}{0}
\setcounter{equation}{0}

\renewcommand{\thefigure}{S\arabic{figure}}
\renewcommand{\theequation}{S.\arabic{equation}}
\renewcommand{\thetable}{S\arabic{table}}

\renewcommand{\thesection}{S\arabic{section}}
\renewcommand{\thesubsection}{\thesection.\alph{subsection}}
\renewcommand{\thesubsubsection}{\thesection.\alph{subsection}.\roman{subsubsection}}

\renewcommand{\thefootnote}{\fnsymbol{footnote}}

%use below for single space
\setlength{\baselineskip}{1.15\baselineskip}  

\newpage
\begin{center}

\textbf{SUPPLEMENTAL MATERIAL}\\
\vspace{2em}
\textbf{\large Revealing quantum geometry effects in magic angle twisted bilayer graphene using the circular photogalvanic effect}\\
\vspace{2em}

\fontsize{9}{12}\selectfont

\vspace{1em}

Eylon Persky,$^{1,2,3*}$ Léonie Parisot,$^{1}$ Minhao He,$^4$ Jiaqi Cai,$^4$ Takashi Taniguchi,$^5$ Kenji Watanabe,$^6$ 
Pierre A. Pantale\'on,$^7$ Francisco Guinea,$^{7,8}$ 
Xiaodong Xu,$^4$ and Aharon Kapitulnik$^{1,2,3,9}$
\bigskip

$^1${\it Geballe Laboratory for Advanced Materials, Stanford University, Stanford, California 94305, USA}\\
$^2${\it Stanford Institute for Materials and Energy Sciences, SLAC National Accelerator Laboratory, 2575 Sand Hill Road, Menlo Park, California 94025, USA}\\
$^3${\it Department of Applied Physics, Stanford University, Stanford, California 94305, USA}\\
$^4${\it Department of Physics, University of Washington, Seattle, Washington, 98195, USA}\\
$^5${\it Research Center for Materials Nanoarchitectonics, National Institute for Materials Science, 1-1 Namiki, Tsukuba 305-0044, Japan.}\\
$^6${\it Research Center for Electronic and Optical Materials, National Institute for Materials Science, 1-1 Namiki, Tsukuba 305-0044, Japan}\\
$^7${\it Imdea Nanoscience, Faraday 9, 28047 Madrid, Spain}
$^8${\it Donostia International Physics Center, Paseo Manuel de Lardiz\'abal 4, 20018 San Sebastián, Spain}
$^9${\it Department of Physics, Stanford University, Stanford, California 94305, USA}\\
\vspace{1em}

\end{center}

\vspace{3em}
\tableofcontents
\vspace{3em}

\section{Experimental Methods}
\vspace{-1mm}
\subsection{Device Fabrication}
\vspace{-3mm}
The heterostructure, which is the same as previously studied and reported in  \cite{He2025,Persky2026}, was assembled using a standard
dry-transfer technique with a PC/PDMS (polycarbonate/polydimethylsiloxane) stamp, and
transferred onto a Si/SiO$_2$ wafer. The twisted bilayer graphene was fabricated by using the tear-and-stack method. The top gate was a 3-5 layer thick graphene flake, in order to minimize its optical absorption. The back graphite gate was 3-5 nm thick. The Hall bar geometry was defined using CHF$_3$/O$_2$ and O$_2$ plasma etching, followed by electron beam lithography. Cr/Au contacts (7nm/70nm) were added using electron beam evaporation. 

\subsection{Optical Setup}
Light from a low-coherence source (superluminescent diode, DenseLight DL-CS5169A) was introduced into the cryostat via a 10 m-long single-mode polarization maintaining fiber (Thorlabs PM-1550XP). A pair of aspheric lenses was used to collimate and re-focuse the light, resulting in a gaussian intensity profile incident on the sample, with a spot size that is determined by the ratio of the two focal lengths. In this study, a spot size of 4.5 $\mu$m was selected to match the dimensions of the device. A zero-order quarter wave-plate (Thorlabs WQP501) was aligned at 45\textdegree\, with respect to the fast axis of the fiber, in order to convert light linearly polarized along the slow/fast axis to right/left circular polarization.

The lens tube was mounted on a piezo-based xy scanner to allow accurate positioning of the beam over the device at low temperatures. A piezo-based stick-slip positioner (Attocube ANPz101) was used to bring the sample into the focal plane of the lens. Finally, a manually-adjustable tip-tilt stage was used at room temperature to align the beam at normal incidence to the sample. At cryogenic temperatures, the lens tube was scanned over the device and the image of the reflected light intensity was compared to the optical microscope image of the device in order to then position the beam over the heterostructure.
\begin{figure*}[ht!]
\centering
\includegraphics[scale=0.5]{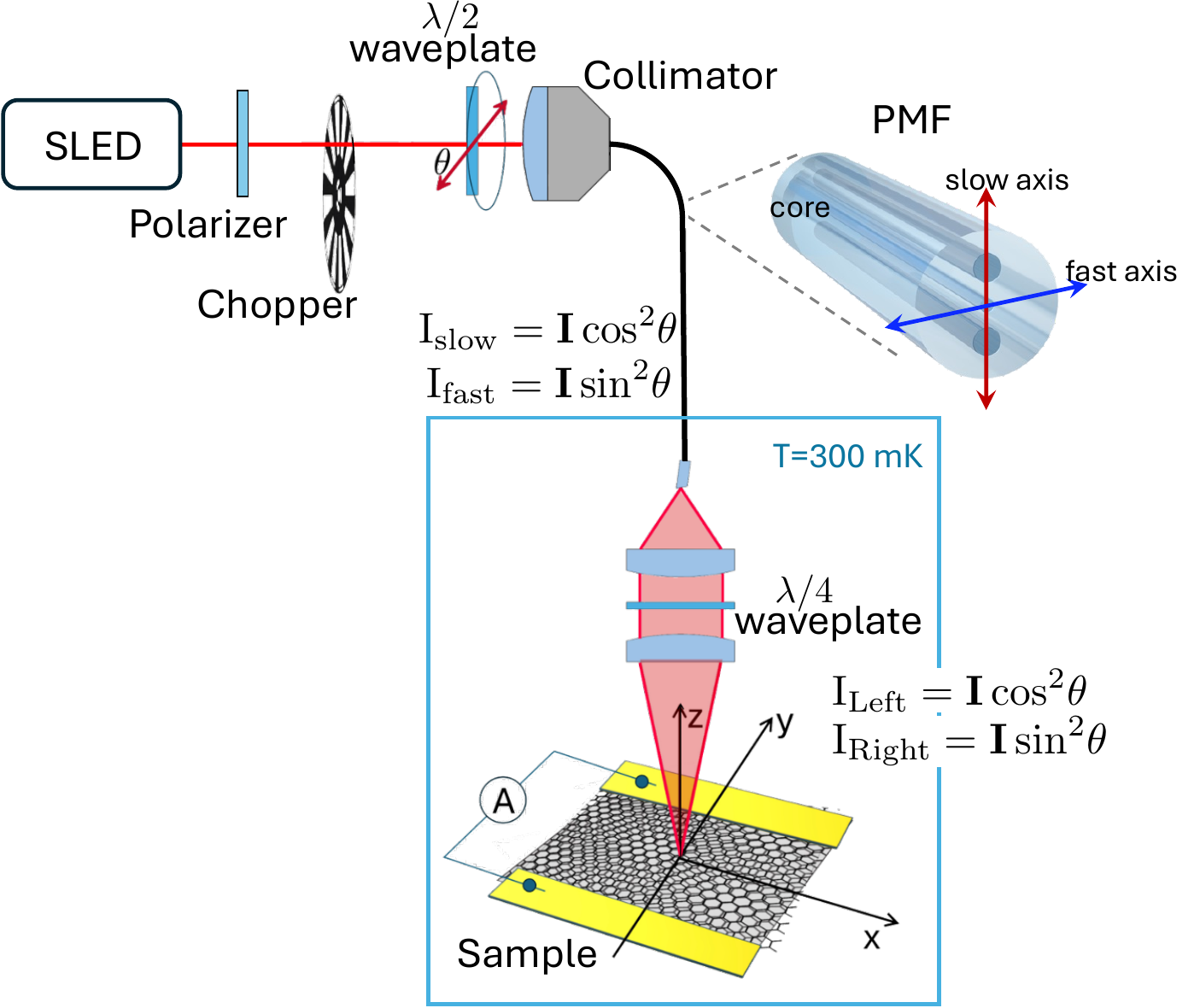}
\captionsetup{justification=centering} 
\caption{Cartoon representation of the experimental setup used in this study. Here a 1550nm SLED source shines through polarizer then a chopper }
\label{setup}
\end{figure*}

\subsection{Transport, Photocurrent and Photovoltage Measurements}
\vspace{-2mm}
The measurements were conducted in a Janis $^3$He cryostat. For transport measurements, an a.c. current (1-5 nA rms, frequency 11 Hz) was applied to the sample, and the resulting voltage was read using a lock-in amplifier. To tune the carrier density and displacement fields, top and bottom gate voltages were applied using two Keithley 2450 units, according to $n = C_T(V_{TG}-V_{TG}^0) + C_B(V_{BG}-V_{BG}^0)$ and $D =  [C_T(V_{TG}-V_{TG}^0) - C_B(V_{BG}-V_{BG}^0)]/2\varepsilon_0$, where $C_T$ ($C_B$) is the capacitance per unit area of the top (bottom) gate. For the photo-current measurement, one electrical contact to the MATBG was grounded and a second contact was connected in series with a trans-impedance amplifier (Keithley 427) and then to ground. A second pair of contacts was connected to a voltage preamplifier (SR560, input impedance 100 M$\Omega$) for simultaneous measurement of the photo-voltage. The reset of the contacts were floating. Lock-in amplifiers were then used to isolate the response at the chopper frequency.

Broadband NIR light (center wavelength- 1550 nm, 3 dB bandwidth, 90 nm) was introduced into a $^3$He cryostat using a 10 m long polarization maintaining fiber and focused onto a 4.5 $\mu$m spot. A quarter wave-plate (QWP), was oriented such that linearly polarized light aligned with the fast (slow) axis of the fiber was converted to left (right) circularly polarized light. Normal incidence was guaranteed by maximizing the back-reflected signal that is coupled back into the fiber. The extreme sensitivity of the back-reflected signal to the tip-tilt ensures that the light is normal to within 0.5 deg. A half wave-plate at room temperature was used to rotate the polarization of the light, adjusting the ratio of left/right circularly polarized light incident on the sample. Note that due to the low coherence of the light, the  proportion of right and left circularly polarized light was changed without having any intermediate linearly polarized light incident on the sample (See Supplemental Material \cite{Supp}). A chopper was used to modulate the intensity of the light, and the photo-current and photo-voltage were measured using lock-in amplifiers. A bottom gate voltage was applied using a Keithley 2450 in order to change the carrier density while the top gate was floating. The resistance of the sample was measured separately by biasing the sample with a small ac current (1-10 nA). Photo-resistivity was also measured by removing the chopper and measuring the resistance under illumination.Photo-resistivity was also measured by removing the chopper and measuring the resistance under illumination.

\subsection{Finite Element Simulations for Photocurrent}
We used a modified Shockley-Ramo theorem \cite{Song2014} to calculate the photocurrent collected at the electrodes, as a function of the beam position. The photocurrent is given by 
\begin{equation}
    I_{ph} = \int d^2r \mathbf{J}_{ph}\cdot \nabla\psi,
\end{equation}
where $\mathbf{J}_{ph}$ is the photo-induced current, and $\psi$ is an auxiliary potential, obtained by solving the Laplace equation $\nabla (\sigma \nabla \psi ) = 0$, where $\sigma$ is the electrical conductivity and the boundary conditions are set such that $\psi = 0$ on the drain electrode and $\psi = 1$ on the source electrode. This ensures that the total current in the device,
$\mathbf{J} = -\sigma\nabla V + \mathbf{J}_{ph}$, satisfies the continuity equation $\nabla\cdot\mathbf{J}_{ph} = 0$.

The photo-induced current has two sources: a thermoelectric effect, where $\mathbf{J}_{th} = -S\sigma \nabla T$, where $S$ is the Seebeck coefficient and $T$ is the temperature, and a photogalvanic contribution, where $\mathbf{J}_{CPGE} \propto  (i\mathbf{E}\times\mathbf{E}^*)_z \mathbf{\hat{x}} $. 

The temperature gradient was estimated by solving a diffusion equation,
\begin{equation}
    -\nabla (\kappa \nabla T) + CT = \frac{P}{2\pi w^2} e^{-(\mathbf{r}-\mathbf{r}_0)^2/2w^2},
\end{equation}
where $\kappa$ is the thermal conductivity of graphene, $C$ is the thermal coupling between the graphene and we assumed the beam has a Gaussian profile with width $w$, a total absorbed power $P$ and centered at $\mathbf{r}_0$.
\vspace{10mm}

\section{Thermoelectric Effects}

\subsection{Experimental Considerations}
\vspace{-2mm}
The Seebeck effect occurs when a temperature difference between two points in an electrically conducting material generates an electromotive force (EMF).  The resulting current density is
\begin{equation}
\textbf{j}=\sigma(-\mathbf{\nabla} V+\textbf{E}_{\rm emf}) \ \ \ \ \ \  {\rm where} \ \ \ \ \ \ \ \textbf{E}_{\rm emf}=-S\mathbf{\nabla} T
\end{equation}

In a typical experiment no current is withdrawn while the voltage generated by the temperature gradient is measured and $S=-\Delta V/\Delta T$. However, in our experiment we measured simultaneously the thermoelectric voltage and the induced current. In such a measurement, with the assumption of a perfect current meter, the measured voltage due to the temperature gradient is
\begin{equation}
\Delta V_{measured}=-S\Delta T +\tfrac{1}{3}\Delta I_{measured} \Delta R
\end{equation}
where R is the resistance and the division by 3 stems from the ratio of distances between the voltage and current leads (see Fig.~2a in the main text). We can estimate that in the metallic regimes between the correlated insulator peaks the finite current yields a correction $\lesssim 1~\mu$V, while at the resistance peaks at $\nu=-3, -2, 0, +1, +2, +3$ the correction is at most 10 $\mu$V.  Comparing these estimates to the thermoelectric photovoltage in Fig.~2d, the thermoelectric photocurrent may introduce at most a 10\% error in the calculated Seebeck coefficient.

\subsection{Guideline Theory}
\vspace{-2mm}
The electronic properties of MATBG are determined by the flat bands and strong correlations, especially near integer fillings.  Let's take the simplest approach where at integer fillings there is a gapped state, while in between we have a metallic state dominated by diffusive scattering.  A simple phenomenological model for the Seebeck coefficient will be
\begin{equation}
S_{\rm tot}(\nu,T)=S_{\rm met}(\nu,T)+S_{\rm gap}(\nu,T)
\label{total}
\end{equation}
Here $S_{\rm met}(\nu,T)$ accounts for the variation of the band structure's density of states as a function of filling excluding the correlated insulating states, and $S_{\rm gap}(\nu,T)$ is the Seebeck coefficient of a gapped insulator with a gap $\Delta(T)$. The simplest approach for a gapped Seebeck coefficient is to say that activated carriers carry energy $\sim\Delta$, so thermovoltage per temperature-K is $\sim \Delta/(eT)$. This gives an estimate:
\begin{equation}
S_{\rm gap}(\nu,T)\sim \pm\frac{k_B}{e}\frac{\Delta}{k_B T}
\end{equation}
where the sign depends if the activated carriers are electrons or holes.\\

The in-between integer fillings will be dominated by the band structure DOS. Here we can use the Mott's formula
\begin{equation}
S(\nu,T)\approx S_{\rm met}(\nu,T) = -\frac{\pi^2 k_B^2 T}{3e}\left.\frac{d\ln\sigma(\epsilon)}{d\epsilon}\right|_{\epsilon=\epsilon_F(\nu)}
\label{guide}
\end{equation}
In these in between metallic regime $S(\nu,T)\propto T$ is expected to be small at low temperatures.

The main issue is how to include a possible energy-dependent scattering rate, which influences the conductivity at the same time as the density of states. Here we propose that in a similar way to treating the optical conductivity of correlated metals, where an  Extended Drude Model is used such that instead of treating the scattering rate $1/\tau$ and mass $m^*$ as constants, they become frequency dependent. This is then parametrized through a fractional power law: $\sigma_1(\omega)\propto \omega^{-\alpha}$ rather than the standard Drude decay $\sigma_1(\omega)\propto \omega^{-2}$ and typically $\alpha <2$. In the same spirit, we can assume that the conductivity that goes into the Mott's formula has the following energy dependence: $\sigma(\epsilon)\propto \epsilon^\alpha$, which yields the approximate Seebeck coefficient:
\begin{equation}
S_{\rm met}(\nu,T) \approx -\frac{\pi^2 k_B^2 T}{3e}\frac{\alpha}{\epsilon_F^*(\nu)}
\end{equation}
With this expression, we see that $S(\nu)$ depends on the sign (electron vs hole), the inverse renormalized Fermi energy $1/\epsilon_F^*(\nu)$, and changes in an effective exponent $\alpha$ near band-structure important features such as van-Hove singularities. This predicts largest thermopower for small $\epsilon_F^*$, often expected near the incompressible fillings. The bottom line is we need to find $\epsilon_F^*(\nu)$ to calculate $S_{\rm met}(\nu,T) $.\\

\section{Circular Photogalvanic Effect: Relevance to our Experiment }
\vspace{-2mm}
\subsection{General Considerations}
\vspace{-2mm}
Applying an electric field $E_j(\omega)=\text{Re}[E_je^{i\omega t}]$,  the transverse current densities up to second order in the electric field are: $J_{j}=\text{Re}\{J^0_{j}+J^\omega_{j}e^{i\omega t}+J^{2\omega}_{j}e^{i2\omega t}\}$. where:
\begin{equation}
J^0_{i}=\chi_{ijl}E_jE_l^*, \ \ \  \ \ J^{\omega}_{i}=\sigma_{ij}E_j \ \ \ \ \ {\rm and} \ \ \ \ J^{2\omega}_{i}=\chi_{ijl}E_jE_l
\end{equation}
Focusing on the rectifying part of the nonlinear current response, we can write
\begin{equation}
J^0_i = J^{0,S}_i +J^{0,A}_i \equiv \zeta_{ijl}(\omega) (E_j E_l^* + E_l E_j^*)+\eta_{ijl}(\omega) (E_j E_l^* - E_l E_j^*)
\end{equation}
where the first term with $\zeta_{ijl}$ describes the symmetric, linear photogalvanic effect and the second term with $\eta_{ijl}$ describes the anti-symmetric, circular photogalvanic effect (CPGE). Focusing on the CPGE response, we define:
\begin{equation}
J^{0,A}_i = \eta_{ijl}(\omega) (E_j E_l^* - E_l E_j^*)\equiv \eta_{ijl}(\omega) \Xi_{jl}, \ \ \ \ \  {\rm where} \ \ 
\qquad \Xi_{jl}=-\Xi_{lj}.
\end{equation}
For circular polarization with $\mathbf{E} = E_0(\hat{x} \pm i\hat{y})$ and at normal incidence light propagates along $z$.
The general form for $\eta_{ijl}$ is:
\begin{equation}
\eta_{ijl}(\omega) = \frac{\pi e^3}{4\hbar^2\omega} \sum_{n,m} \int \frac{d^2k}{(2\pi)^2} f_{nm}(\mathbf{k}) \, \mathcal{T}_{ijl}^{nm}(\mathbf{k}) \, \delta(\omega_{mn}(\textbf{k}) - \omega)
\end{equation}
where in general we include lifetime broadening through the replacement:
\begin{equation}
\delta(\omega_{mn}-\omega)\ \longrightarrow\ \delta_\Gamma(\omega_{mn}-\omega) \equiv \frac{1}{\pi}\frac{\Gamma}{(\omega_{mn}-\omega)^2+\Gamma^2}
\end{equation}
Within this formulation the injection tensor contains three contributions:
\begin{equation}
\mathcal{T}_{ijl}^{nm} = \mathcal{T}_{ijl}^{nm,\text{I}} +  \mathcal{T}_{ijl}^{nm,\text{shift}} + \mathcal{T}_{ijl}^{nm,\text{BC}} 
\end{equation}
where the Standard Injection Current (Asymmetric Scattering) is:
\begin{equation}
\mathcal{T}_{ijl}^{nm,\text{I}} = 2\, \text{Im}[r_{nm}^j r_{mn}^l] \, v_{mn}^i
\end{equation}
where $v_{mn}^i = \frac{1}{\hbar}\partial_{k_i}(E_m - E_n) = v_m^i - v_n^i$ is the velocity difference and $r_{nm}^i = \bra{n}i\partial_{k_i}\ket{m}$ is the interband position matrix element.\\

The shift current is the other principal interband photocurrent, which is a major contribution in linear photogalvanic effect (LPGE). In general
\begin{equation}
\mathcal{T}_{ijl}^{nm,\text{shift}} = 2\, \text{Im}[r_{nm}^j r_{mn}^l] \, R_{nm}^i
\end{equation}
where the shift vector is: $R_{nm}^i(\mathbf{k}) = i\sum_{p \neq n,m} \left(\frac{r_{np}^i r_{pm}^l}{\omega_{pn}} - \frac{r_{np}^l r_{pm}^i}{\omega_{pm}}\right) + \partial_{k_i}\arg[r_{nm}^l]$.

However, careful analysis of this expression when calculated for the configuration of normal-incidence CPGE shows that circular light eliminates the anisotropic part of the shift current, with the only surviving piece originates from 
\begin{equation}
j_i^{\text{shift}}\big|_{\text{circ}}=\frac{|E_0|^2}{2}
\big(\sigma_{ixx}^{\text{shift}}+\sigma_{iyy}^{\text{shift}}\big).
\end{equation}

For crystals with $\ge3$-fold rotation ($C_3,C_{3v},C_4,C_6,\dots$) the in-plane trace vanishes and circular light gives no shift current. In our experiments, CPGE is measured through the difference in current between right and left circularly polarized lights and thus the shift current, being helicity-even, cancels exactly in this difference, regardless of symmetry.

\subsection{Berry Curvature Effects}
\vspace{-2mm}
In a standard approach we would expect that in equilibrium we could distinguish  interband from itnraband effects:
\begin{equation}
\mathcal{T}_{ijl}^{nm,\text{BC}}(\omega)=\mathcal{T}_{ijl}^{nm,\text{BC-inter}} +\mathcal{T}_{ijl}^{nn,\text{BC-intra}}
\end{equation}
with
\begin{equation}
\mathcal{T}_{ijl}^{nm,\text{BC-inter}} =2\, \text{Im}[r_{nm}^j r_{mn}^l] \, \Delta\Omega_{nm}^{i}
\end{equation}
where $\Delta\Omega_{nm}^{i} = \Omega_m^i(\mathbf{k}) - \Omega_n^i(\mathbf{k})$ is the Berry curvature difference between bands $m$ and $n$.\\

As for the intraband contribution, reverting to $\eta_{ijl}$, we expect a direct dependence on the Berry curvature dipole tensor:
\begin{equation}
\eta^{\mathrm{BC\text{-}intra}}_{ijl}(\omega)=\frac{e^{3}\tau}{2\hbar^{2}}\,\frac{1}{1+\omega^{2}\tau^{2}}\,\Big(\epsilon_{ijp}D_{pl}-\epsilon_{ilp}D_{pj}\Big)
\end{equation}
where $\epsilon_{ijp}$ is the totally antisymmetric Levi-Civita tensor, and $D_{pl}$ is the intraband Berry curvature dipole given by:
\begin{equation}
D_{pl}=\sum_{n}\int\frac{d^2k}{(2\pi)^2}\, f_{n\mathbf k}\,\partial_{k_l}\Omega_{n}^{p}(\mathbf k)\;=\;-\sum_{n}\int\frac{d^2k}{(2\pi)^2}\,(\partial_{k_l} f_{n\mathbf k})\,\Omega_{n}^{p}(\mathbf k)
\label{pierre}
\end{equation}
Note that in strictly 2D only $\Omega_n^z$ is nonzero and thus only $D_{zl}$ is relevant.  Obviously, for high frequencies such as our excitation light at $\hbar\omega\approx 0.8$ eV, we expect $\omega\tau\gg1$ and therefore a strong reduction in capturing this effect.

\subsection{Berry Curvature Dipole from Optical Pumping}
\vspace{-3mm}
However, based on the remarkable correspondence between our CPGE data and calculations that focus on the Fermi-surface using eqn.~\ref{pierre}, we note that the above intraband approach can also emerge through an optical pumping process, where the high-energy photon excites carriers via interband transitions $n\leftrightarrow m$, which in turn creates a nonequilibrium $\delta f_n(\mathbf k;\omega)$ or $\delta f_m(\mathbf k;\omega)$.  Since our aim is to focus on pure CPGE at normal incidence, we use the superscript $(A)$ to select only the helicity-odd optical pumping kernel and focus on 2D
\begin{equation}
\mathcal W^{(A)}_{nm,jl}(\mathbf k;\omega)=\frac{2\pi e^2}{\hbar^2}\,\text{Im}[r_{nm}^j r_{mn}^l]\,\delta_\Gamma(\omega_{mn}-\omega)
\end{equation}
The respective helicity-odd change in distribution function is:
\begin{equation}
\delta f_{n\mathbf k}^{(A)}(\omega)=\tau_{n}^{\rm rec}\;\sum_{m\neq n}f_{nm}(\mathbf k)\;\mathcal W^{(A)}_{nm,jl}(\mathbf k;\omega)\;\Xi_{jl}
\end{equation}
where $f_{nm}=f_n-f_m$ and $\tau_n^{\rm rec}$ is a phenomenological population relaxation/recombination time for the photoexcited distribution (can differ strongly from momentum relaxation time).\\

By analogy to a current induced by a finite magnetic field, applying a monochromatic electric field $\textbf{E}(t)={\rm Re}\{\textbf{E}e^{-i\omega t}\}$ in the presence of non-trivial quantum geometry contribution \cite{Yu2025}, to the band structure, the modified velocity operator for band $n$  ( $\textbf{B} = 0$) acquires an anomalous term:
\begin{equation}
\dot{\textbf{r}}=\textbf{v}_n(\textbf{k})-\frac{e}{\hbar}\,\textbf{E}\times \bm\Omega_n(\textbf{k}),
\qquad
\hbar\dot{\textbf{k}}=-e\textbf{E}.
\label{anomvel}
\end{equation}
The anomalous velocity term yields an anomalous current
\begin{equation}
\textbf{J}^{\rm an}(\textbf{r},t)=-e\sum_n\int\frac{d^dk}{(2\pi)^d}\,\left(-\frac{e}{\hbar}\textbf{E}(\textbf{r},t)\times \bm\Omega_n(\textbf{k})\right)\,f_n(\textbf{r},\textbf{k},t).
\label{eqnjan0}
\end{equation}
Photogalvanic currents emerge where spatially uniform illumination is applied to a sample. In that case the Boltzmann equation in the relaxation time approximation is
\begin{equation}
\frac{\partial f_n}{\partial t} + \dot{\textbf{k}}\cdot \nabla_{\textbf{k}} f_n = -\frac{f_n - f_n^0}{\tau},
\label{eqnboltzmann}
\end{equation}
where $f_n^0(\textbf{k})=f_\text{FD}(\epsilon_n(\textbf{k}))$ is the equilibrium Fermi--Dirac distribution.  Expanding $f_n$ in powers of the field, the first correction term
\begin{equation}
f_n(\textbf{k},t)=f_n^0(\textbf{k}) + f_n^{(1)}(\textbf{k},t)+f_n^{(2)}(\textbf{k},t)+\cdots,
\end{equation}
we are interested in the helicity-odd part of the first correction, where we identify: $f_n^{(1)}(\textbf{k},t)\equiv \delta f_{n\mathbf k}^{(A)}(\omega)$, which here it corresponds to a ``pumped Berry curvature current'' given by (in 2D): 
\begin{equation}
J_i^{\rm (BC,\,pumped)}(\omega)
=
-\frac{e^2}{\hbar}\,
\sum_n \int\frac{d^2k}{(2\pi)^2}
\epsilon_{ijz}\,\Omega_n^z(\mathbf k)\;
\mathcal D_n(\omega)\;
\Big[ \delta f^{(A)}_{n\mathbf k}(\omega)\Big]\;
E_j
\end{equation}
where we added a Drude frequency factor $\mathcal D_n(\omega)=1/(1+i\omega\tau)$.\\

In a similar fashion that we obtained the Berry curvature dipole in the equilibrium case, here to obtain the ``pumped Berry curvature dipole'' we need to expand the current to the next order derivative of the nonequilibrium distribution. 
\begin{equation}
D^{zl}_{\rm pumped}(\omega)
\equiv
\sum_n \int\frac{d^2k}{(2\pi)^2}
\partial_{k_l}\Omega_n^z(\mathbf k)\;
\delta f^{(A)}_{n\mathbf k}(\omega)
\end{equation}
Substituting $\delta f^{(A)}$ we obtain the final result:
\begin{equation}
D^{zl}_{\rm pumped}(\omega)
=
\sum_n \tau_n^{\rm rec}\int\frac{d^2k}{(2\pi)^2}
\partial_{k_l}\Omega_n^z(\mathbf k)\;
\sum_{m\neq n}
f_{nm}(\mathbf k)\;
\mathcal W^{(A)}_{nm,jl}(\mathbf k;\omega)\;
\Xi_{xy}
\end{equation}

\subsection{The ``Pumped'' BCD and Application to Optically-Pumped BCD in MATBG}
\vspace{-2mm}

In the manuscript we argue that the Berry curvature dipole that gives rise to the CPGE current in our experiments seems to follow the equilibrium BCD at the Fermi surface, which is dominated by the flat band. Therefore, we first ask the question of what are the conditions that the flat band dominates the behavior, at least in a proportionality fashion.

 For this to happen, we first require that the thermalization time scale, $\tau_{\rm th}$ is much shorter compared to either momentum relaxation or recombination time, so that the band reaches a quasi-Fermi distribution. That is, after a fast cascade into the flat bands and rapid intraband thermalization $\delta f^{(A)}$ can become quasi-thermal , which means that $D_{\rm pumped}$ can become $\propto D_{\rm eq}$ (where the subscript ``eq'' stands for the equilibrium BCD).  In particular, we require that light doesn’t activate a new symmetry channel that equilibrium doesn’t have.  Within such conditions 
$\delta f_{n\mathbf{k}}^{(A)}$ is practically a  correction to the equilibrium distribution function in the flat band. Focusing on the flat band, with its equilibrium distribution $f_F^{(0)}$, we can write
\begin{equation}
\delta f_{F\mathbf {k}}\approx
-\frac{\partial f^{(0)}_F}{\partial \epsilon}\,\delta\mu
-\frac{\partial f^{(0)}_F}{\partial T}\,\delta T,
\end{equation}
and in thermal equilibrium, only the chemical-potential shift piece survives, giving exactly the equilibrium BCD weighting
\begin{equation}
D^{zl}_{\rm pumped}\approx \delta\mu \int_{\mathbf k}\left(-\frac{\partial f_F^{(0)}}{\partial\epsilon}\right)\partial_{k_l}\Omega^z
= \delta\mu \, D^{zl}_{\rm eq}
\end{equation}
Note also that we also assume that during the illumination process the Berry curvature landscape is the equilibrium one:   $\Omega_f^z(\mathbf k)\text{(during illumination)}\approx\Omega_{f,{\rm eq}}^z(\mathbf k).$ This is justified in our experiments where we demonstrated no discernible photodoping effects with the light intensity we used \cite{Persky2026}.

Other routes to obtaining BCD that is roughly proportional to the equilibrium flat bands involve a pumping kernel that is almost \textbf{k}-independent. However, at $\hbar\omega\sim0.8$ eV in MATBG, resonant transitions usually involve remote bands with sizable dispersion; without strong broadening/relaxation, the pumping is typically quite $\mathbf k$-selective.\\

With the above assumptions, focusing on optically-pumped scenario, we indeed expect that  
\begin{equation}
D^{zl}_{\rm pumped}(\omega)\approx C(\omega)\,D^{z\l}_{\rm eq}
\end{equation}
where $C(\omega)$ is a proportionality factor that depends on the light frequency. At least on the side of negative fillings this scenario seems well justified. 
\vspace{10mm}

\section{The Hartree-Renormalized Strained MATBG}
\vspace{-1mm}
\subsection{General Remarks}
\vspace{-3mm}
In our calculations the electron-electron interaction is introduced with a self-consistent Hartree potential in the TBG/hBN Hamiltonian where to the usual strained TBG Hamiltonian \cite{San-Jose2014}, a term that captures the effect of hBN on the adjacent graphene layer  \cite{Cea2020} and a Hartree potential are added.  Specifically, for each filling $\nu$, the single-particle Hamiltonian is taken as
\begin{equation}
H_\nu(\textbf{k})=H_0(\textbf{k})+V_{\text{H}}(\nu)
\end{equation}
Here $H_0(\textbf{k})=H_{\text{TBG}}+V_{\text{hBN}}$ is the strained TBG Hamiltonian, including the chosen continuum-model terms, and $V_{\text{H}}(\nu)$ is the static Hartree potential at filling $\nu$ \cite{Cea2019}. We assume that $\nu=0$ is defined at half filling of the system. The eigenvalue problem is
\begin{equation}
H_\nu(\textbf{k})|u_{n\textbf{k}}^{(\nu)}\rangle=\epsilon_{n\textbf{k}}^{(\nu)} |u_{n\textbf{k}}^{(\nu)}\rangle
\end{equation}
and the respective band velocities are calculated from:
\begin{equation}
v_{n\textbf{k}}^i=\frac{1}{\hbar}\langle u_{n\textbf{k}}^{(\nu)}| \partial k_i H_\nu(\textbf{k})|u_{n\textbf{k}}^{(\nu)}\rangle
\end{equation}
where $i=x,y$. Note that the Hartree potential is independent of $\textbf{k}$, then it does not contribute directly to $\partial k_i H_\nu$, but it still changes the velocities through the Hartree-renormalized eigenvectors.

\subsection{BCD Calculations}
\vspace{-3mm}
Applying an electric field $E_j(\omega)=\text{Re}[E_je^{i\omega t}]$,  the transverse current densities up to second order in the electric field are: $J_{j}=\text{Re}\{J^0_{j}+J^\omega_{j}e^{i\omega t}+J^{2\omega}_{j}e^{i2\omega t}\}$. where:
\begin{equation}
J^0_{i}=\chi_{ijl}E_jE_l^*, \ \ \  \ \ J^{\omega}_{i}=\sigma_{ij}E_j \ \ \ \ \ {\rm and} \ \ \ \ J^{2\omega}_{i}=\chi_{ijl}E_jE_l
\end{equation}
The nonlinear susceptibility can be written in terms of the Berry curvature dipole $D_{pl}$ (see Supplemental Material \cite{Supp})
\begin{equation}
\chi_{ijl}=\frac{e^3\tau}{2\hbar^2}\epsilon_{ijp}D_{pl}
\end{equation}
where
\begin{equation}
D_{pl}=-\sum_n \int\frac{d^2k}{(2\pi)^2}
\Omega_n^p(\mathbf k)\;
\partial_{k_l}f^0_{n\textbf{k}}
\end{equation}
Here $f^0_{n\textbf{k}}$ is the equilibrium Fermi-Dirac distribution of band $n$ and $\tau$ is the relevant scattering time.  The calculations proceed as prescribed in Ref.~\cite{Pantaleon2022}, where the wavefunctions required to obtain the Berry curvature are calculated self-consistently for each value of the Fermi energy.  The cBCD alculations are performed in the limit $\omega\tau\ll 1$ as appropriate for low frequency light excitation. As explained in the text, this is justified within the assumption of ``pumped BCD mechanism.'' Since the experiments are performed at normal incidence, the relevant components of the CPGE tensor are $\eta_{xxy}=-\eta_{yxx}$ and $\eta_{yyx}=-\eta_{xyy}$. Define an angle $\gamma$ between the BCD-direction and the $x$-axis (we identify $x \equiv  \textbf{a}_1$ the unit cell primitive vector), the effective angular BCD is 
\begin{equation}
D_{eff}(\gamma)=D_{xz}{\rm cos}\gamma+D_{yz}{\rm sin}\gamma
\end{equation}
with results shown in Fig.~\ref{cpgeX}(b).

\begin{figure}[ht!]
\centering
\includegraphics[width=0.7 \columnwidth]{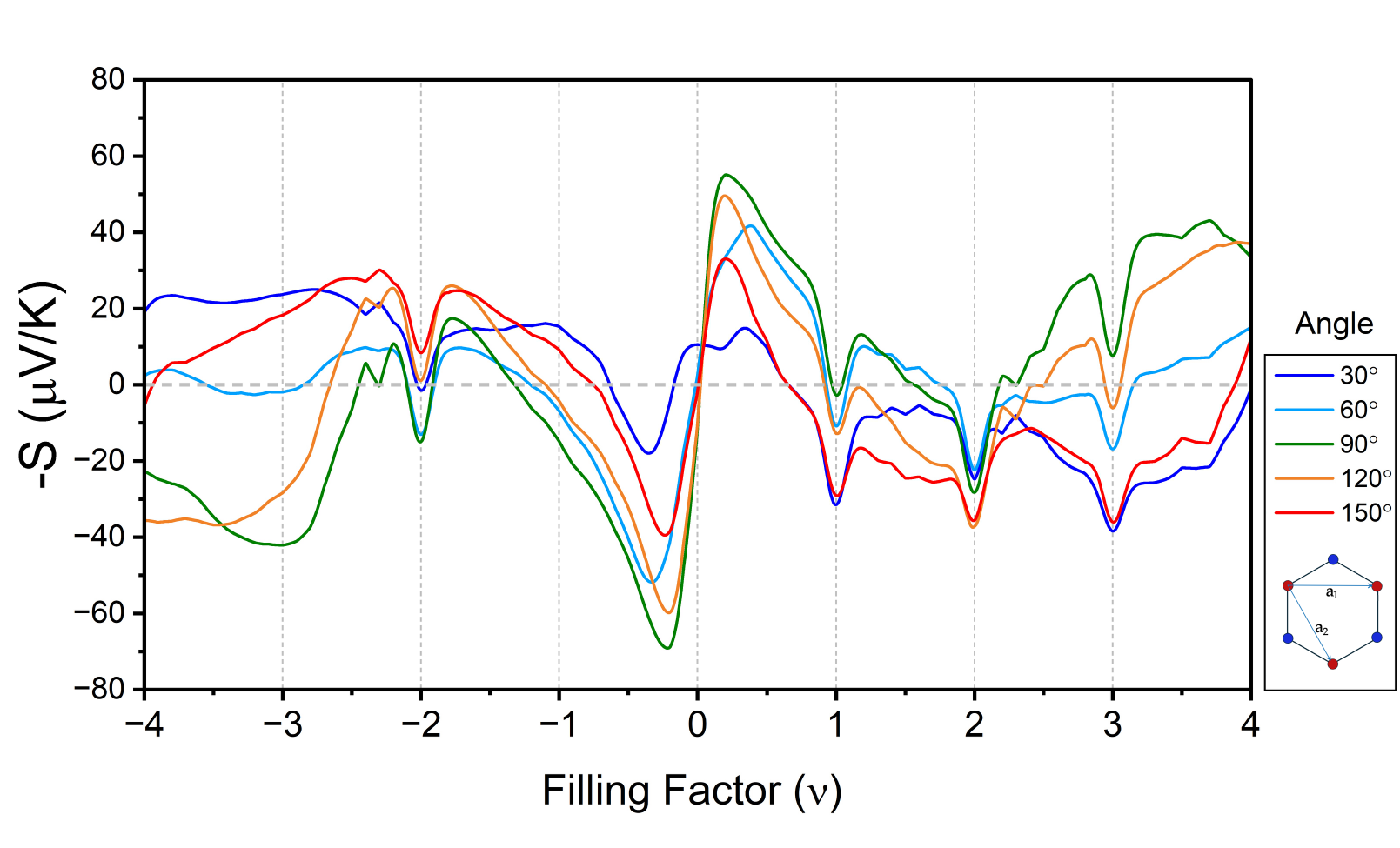}
\caption{ Filling dependent Seebeck coefficient for TBG with a phenomenological integer-filling gap correction with the same parameters as in Fig. 1b. Each color corresponds to the different angular direction $\gamma$ with respect to the graphene vector $\textbf{a}_1$. }
\label{teppp}
\end{figure}

To compare with our CPGE results, If Hartree is the main filling-dependent effect and correlations are weak (excluding the sharp features at integer fillings), we may expect broad features associated with the van-Hove singularities (vHs). To compare to our photothermoelectric voltage, guiding by Eqn.~\ref{guide}, we plot in Fig.~\ref{teppp}  the negative of the Seebeck coefficient vs. the moir\'e band filling factor.  The calculation follows the guideline of Eqn.~\ref{total} with the temperature gradient pointing in different angles to a graphene axis. The temperature is assumed to be 10 K and the parameters are the same as for the BCD calculation in Fig.1(b).  While the experimental results of photovoltage were not done at zero-current configuration, the measured photovoltage (Fig.2(d)), particularly for orientation $\gamma$ between $90^\circ$ and $120^\circ$, is captured well by the calculations particularly around the neutrality point, suggesting that the MATBG+substrate+strain+interactions at the Hartree level can account for the overall behavior. Adding the correlated states at integer fillings ``by hand'' points for the need of further considerations.

\subsection{Seebeck Coefficient Calculations}
\vspace{-3mm}
To calculate the Seebeck coefficient, we first note that the Fermi-window at a given temperature $T$ is given by
\begin{equation}
-\frac{\partial f}{\partial \epsilon}=\frac{1}{4k_BT} {\rm sech}^2 \left( \frac{\epsilon-\mu}{2k_B T} \right)
\end{equation}
where $\mu$ is the chemical potential. The conductivity moment is
\begin{equation}
L^0_{ij}=\frac{1}{\mathcal{N}}\sum_{n,\textbf{k}}w_{\textbf{k}}\mathcal{V}_{ij}(n,\textbf{k})\left(-\frac{\partial f}{\partial \epsilon}\right)_{\epsilon=\epsilon_{n\textbf{k}}^{(\nu)}}
\end{equation}
and the thermoelectric moment
\begin{equation}
L^1_{ij}=\frac{1}{\mathcal{N}}\sum_{n,\textbf{k}}w_{\textbf{k}}\left(\epsilon_{n\textbf{k}}^{(\nu)}-\mu\right)\mathcal{V}_{ij}(n,\textbf{k})\left(-\frac{\partial f}{\partial \epsilon}\right)_{\epsilon=\epsilon_{n\textbf{k}}^{(\nu)}}
\end{equation}
Here $\mathcal{N}=\sum_{\textbf{k}} w_k$ is the normalization, which we will take with equal weight $w_{\textbf{k}}=1$ and the velocity weights are $\mathcal{V}_{ij}=v_{n\textbf{k}}^i(v_{n\textbf{k}}^{j})^*$.

Define as above the angle $\gamma$ between the temperature gradient and the $x$-axis (we identify $x \equiv  \textbf{a}_1$ the unit cell primitive vector), the Seebeck coefficient is calculated from
\begin{equation}
-S_\gamma=\frac{1}{T}\frac{L^1_{\gamma\gamma}}{L^0_{\gamma\gamma}}
\end{equation}\\
where $L^a_{\gamma\gamma}\equiv L^a_{xx}{\rm cos}^2\gamma+ L^a_{yy}{\rm sin}^2\gamma+2L^a_{xy}{\rm cos}\gamma{\rm sin}\gamma$, $a=0,1$, thus the moments must be combined first, and only then should the ratio be taken.\\

\end{document}